\documentclass[pre,preprint,showpacs,preprintnumbers,amsmath,amssymb]{revtex4}
\usepackage{graphics,epstopdf,epsfig,stmaryrd}

\newcommand{\dw}{\delta\omega_p}

\newcommand{\ea}{\textit{et al.}}

\begin {document}
\title{Nonlinear kinetic modeling of stimulated Raman scattering in a multidimensional geometry} \author{D. B\'enisti}
\email{didier.benisti@cea.fr} \author {O. Morice}
\author{L. Gremillet} \author{A. Friou} \author{E. Lefebvre}

\affiliation{ CEA, DAM, DIF F-91297 Arpajon, France.} \date{\today}
\begin{abstract}
In this paper, we derive coupled envelope equations modeling the growth of stimulated Raman scattering (SRS) in a multi-dimensional geometry, and accounting for nonlinear kinetic effects. In particular, our envelope equations allow for the nonlinear reduction of the Landau damping rate, whose decrease with the plasma wave amplitude depends on the rate of side-loss. Account is also made of the variations in the extent of the plasma wave packet entailed by the collisionless dissipation due to trapping. The dephasing between the electron plasma wave (EPW) and the laser drive,  as well as the self-focussing of the plasma wave, both induced by the EPW nonlinear frequency shift, are also included in our envelope equations. These equations are solved in a multi-dimensional geometry using our code dubbed BRAMA, whose predictions regarding the evolution of Raman reflectivity as a function of the laser intensity are compared against previously published PIC results, thus illustrating the ability of BRAMA simulations to provide the correct laser threshold intensity for SRS, as well as the right order of magnitude of Raman reflectivity above threshold.
\end{abstract}
   
\maketitle
\section{Introduction} 
\label{I}
The present work is part of the effort that has been undertaken for about one decade by several groups in order to understand, and efficiently model, the impact of nonlinear kinetic effects on the growth of stimulated Raman scattering (SRS). All the recent modelings we know of, including the one presented here, rely on the hypothesis that the wave electric fields may be written in terms of a slowly varying envelope and a rapid phase. Using this approximation, Vu~\ea~developed a  reduced Particle In Cell (PIC) code, described in~Ref.~\cite{vu99}~(and a further simplified PIC-like scheme was even derived by Hur~\ea~in Ref.~\cite{hur04}), which they used in~Ref.~\cite{vu01}~to show that Raman reflectivity could reach much higher levels than linear theory would predict. This so-called ``kinetic inflation'' was confirmed experimentally by Montgomery~\ea~in Ref.~\cite{montgomery},~as well as by full kinetic simulations (see for example Refs.~\cite{strozzi,yin08}), and was attributed to the nonlinear reduction of the Landau damping rate of the electron plasma wave (EPW), following the original idea of O'Neil~\cite{oneil}. Such results triggered several theoretical developments, for example those of Refs.~\cite{rose01,lindberg08, yampo, benisti07, benisti09}, revisiting O'Neil's work and aiming at a formulation of the collisionless damping rate general enough to apply to a realistic physics situation,  such as SRS. 

Another important nonlinear kinetic effect, widely discussed this last decade, is the nonlinear frequency shift of the EPW, which may induce a dephasing between the laser drive and the plasma wave and, therefore, hamper the growth of SRS, as argued for example in~Ref.~\cite{vu01}. Moreover, in more than one dimension (1-D), the frequency shift usually has a transverse profile making the EPW phase velocity smaller close to the wave axis of propagation than away from it, causing the so-called ``wave front bowing" evidenced numerically in Ref.~\cite{yin07}. In practice, this entails a transverse component in the plasma wave number that tends to make the EPW self-focus. Just like for the nonlinear collisionless damping rate, several recent theoretical papers (for example Refs.~\cite{rose01,benisti08,lindberg}) revisited the original works on the nonlinear frequency shift of an EPW, Refs.~\cite{morales,dewar}.

The aforementioned theoretical works were mainly motivated by the will to  develop envelope codes running fast enough to address such a large scale system as a fusion hohlraum \cite{lindl}, and accurately accounting for nonlinear kinetic effects. Such envelope codes were described in Refs.~\cite{vu07,yampo, lindberg08, brama}~and their predictions were compared with those of 1-D kinetic simulations. In this paper, we go further in that direction, and show what we believe are unprecedented comparisons between results from PIC simulations and from our envelope code, BRAMA, for Raman reflectivity in a two-dimensional (2-D) and a three-dimensional (3-D) geometry.

In BRAMA, the modeling of SRS rests on the hypotheses that the plasma is collisionless and non relativistic, and that the waves are purely sinusoidal. The non-relativistic approximation is clearly valid for the laser intensities used in our simulations. Moreover, for the range of parameters considered in this paper, the main effect of collisions, which is to partially restore Landau damping (see Ref. \cite{vu07} for details), is overcome by  that due to the transverse electron detrapping. The sinusoidal assumption is vindicated \textit{a posteriori}~by our results showing that SRS saturates at EPW amplitudes small enough for the effects of  anharmonicity to be negligible. This allows us to use in BRAMA the values derived in Ref.~\cite{benisti08} for the frequency shift, $\dw$, which are in very good agreement with those inferred from 1-D Vlasov simulations of SRS, and which rest of the hypothesis that the electron motion is nearly adiabatic.  The latter approximation is valid when $l_\bot \agt \lambda_l$, where $l_\bot$ is the transverse extent of the plasma wave and $\lambda_l$ the laser wavelength (see Ref.~\cite{benisti07} and Appendix \ref{A}), a condition we were careful to meet in our BRAMA simulations. The dephasing between the plasma wave and the laser drive induced by $\dw$ is accounted for within the hypothesis of self-optimization, which is recalled in Section \ref{III} , and which also is supported by 1-D Vlasov results (see Ref.~\cite{PRL_10}). As for the self-focussing induced by $\dw$, it is explicitly contained in our envelope equations. 

The nonlinear expression we use for the EPW collisionless damping rate follows from the theoretical developments of Ref.~\cite{benisti09}, and allows for transverse electron trapping and detrapping. This value for the damping rate was checked in Ref.~\cite{benisti09}~to agree with results obtained from 1-D Vlasov simulations of SRS,  and was shown in Ref.~\cite{compare}~to be consistent with that of Yampolsky and Fisch~\cite{yampo}~over a finite range of wave amplitudes. In BRAMA, we also account for the collisionless dissipation of the plasma wave induced by trapping, which entails no damping but alters the extent of the plasma wave packet (both in the longitudinal and transverse directions) due to a nonlinear and non local change in the EPW group velocity, whose theoretical estimate was checked in Refs.~\cite{benisti09,vgroup} to be in very good agreement with results from 1-D Vlasov simulations of SRS. 

We are therefore using in this paper a nonlinear kinetic modeling of SRS which we checked very carefully using 1-D Vlasov simulations. However, other nonlinear kinetic effects than those previously cited exist, for example the growth of sidebands evidenced in Refs.~\cite{brunner,yin08,rousseaux}, and may have important implications in SRS. These are nevertheless not accounted for in BRAMA, and, in spite of the imperfections of our modeling, we find it interesting to discuss its ability to make relevant predictions as regards Raman reflectivity. This is the purpose of the present paper to do so by showing comparisons between BRAMA results and the PIC simulation results obtained by Yin~\ea~in Refs.~\cite{yin08,yin09}. 

This article is organized as follows. In the next Section, we derive the envelope equations solved in BRAMA, and discuss them physically. Section \ref{III} is devoted to the comparisons we made between the results from PIC and BRAMA simulations regarding the evolution of Raman reflectivity as a function of the laser intensity. Section \ref{IV} summarizes and concludes this paper. Moreover, supplementary material and discussions are given in the Appendices.

\section{Wave equations}
\label{II}
In this paper, Raman scattering is studied within the context of the so-called three-wave model, wherein the total electric field reads
\begin{equation}
\label{1}
\vec{E}_{tot}=-i \hat{x}_p\frac{\mathcal{E}_p}{2}e^{i(\varphi_p^{\text{lin}}+\delta \varphi_p)}-i \hat{y}_l\frac{\mathcal{E}_l}{2}e^{i\varphi_l^{\text{lin}}}+\hat{y}_s\frac{\mathcal{E}_s}{2}e^{i\varphi_s^{\text{lin}}}+c.c.,
\end{equation}
where $\varphi_w^{\text{lin}}=k^{\text{lin}}_w x - \omega^{\text{lin}}_w t$, with $w=p,l,s$ respectively for the plasma, laser and scattered wave, and $\varphi_l^{\text{lin}}=\varphi_p^{\text{lin}}+\varphi_s^{\text{lin}}$. Moreover, $(\omega^{\text{lin}}_{l,s} )^2=\omega_{pe}^2+(k_{l,s}^{\text{lin}}c)^2$, $c$ being the speed of light in vacuum and $\omega_{pe}$ the plasma frequency. $\omega^{\text{lin}}_{l}$ is the vacuum laser frequency, while $k^{\text{lin}}_{p}$ and $\omega^{\text{lin}}_{p}$ are such that they maximize the linear SRS growth rate derived in Ref.~\cite{drake}. As for $\delta \varphi_p$, it accounts for the effect of the frequency shift, $\dw=-\partial_t \delta \varphi_p$, while the $\mathcal{E}_w$'s are slowly varying envelopes, $\vert \mathcal{E}_w^{-1}\partial_x\mathcal{E}_w \vert \ll \vert k_w^{\text{lin}} \vert$ and $\vert \mathcal{E}_w^{-1}\partial_t\mathcal{E}_w \vert \ll \vert \omega_w^{\text{lin}} \vert$. Moreover, as shown in Appendix \ref{A}, the polarization $\hat{x}_p$ is very close to the direction of the wave number, $\vec{k}_p\equiv k_p^{\text{lin}} \hat{x}+\vec{\nabla} \delta \varphi_p$, because the plasma wave is nearly electrostatic. As for the laser and scattered waves, we account for the fact that they are not exactly polarized along the $\hat{y}$ direction, and derive in Appendix~\ref{A} the relation between their $x$ and $y$ components. 

\subsection{Equation for the plasma wave}
In order to derive the envelope equation for the plasma wave amplitude, we use the same procedure as in Ref.~\cite{dissipation} which consists in first using the results from Whitham's variational approach~\cite{whitham}, and in then adding~``by hand" the effects of dissipation and of the electromagnetic drive, which are not accounted for in Whitham's theory. This procedure can be justified \textit{a posteriori} by going through Vlasov-Gauss equations as was done in the Appendix of Ref.~\cite{vgroup}, or in Ref.~\cite{benisti07}. This program will however not be pursued here, and we prefer refer the reader to Ref.~\cite{dissipation}~where our procedure appears quite naturally. 

Hence, we first start by using Whitham's variational approach, which yields 
\begin{equation}
\label{B2}
\frac{\partial^2 L}{\partial t \partial \omega_p}-\vec{\nabla}.\left[\vec{\nabla}_{\vec{k}_p}L \right]=0,
\end{equation}
where $\omega_p\equiv \omega_p^{\text{lin}}-\partial_t \delta \varphi_p$ is the nonlinear frequency of the plasma wave. We henceforth assume that $\mathcal{E}_p$ is real so that the Lagrangian density is $L= \int_0^{\mathcal{E}_p} \partial_{\mathcal{E}'}L d\mathcal{E}'$, where $\partial_{\mathcal{E}'}L=0$ is the dispersion relation of a freely propagating EPW,  at 0-order in the variations of $\mathcal{E}_p$. At this order, the plasma wave may clearly be considered electrostatic, so that $\partial_{\mathcal{E}'}L=(1+\chi_r)\mathcal{E}'$, where $\chi_r$ is the adiabatic approximation of the real part of the electron susceptibility. It  assumes the same values as in 1-D, derived in~Ref.~\cite{benisti07}, and only depends on the plasma wave number through  its modulus, $k_p$, since the EPW may be considered electrostatic and the electron distribution function is isotropic. Then, using  $L\equiv \int_0^{\mathcal{E}_p} \left[1+\chi_r\mathcal{E}'\right]d\mathcal{E}'$ and the consistency relation $\partial_t \vec{k}_p=-\vec{\nabla}\omega_p$, Eq.~(\ref{B2}) yields,
\begin{equation}
\label{B3} 
\frac{\partial \chi_r}{\partial \omega_p} \frac{\partial \mathcal{E}_p}{\partial t}-\frac{\partial \chi_r}{\partial k_p}\frac{\partial  \mathcal{E}_p} {\partial x_p}+\frac{\mathcal{E}_p}{2} \left[Ê\frac{\partial \omega_p}{\partial t}\frac{\partial ^2\chi_r}{\partial \omega_p^2}-\vec{\nabla}.\hat{x}_p  \frac{\partial \chi_r}{\partial k_p}- \hat{x}_p.\vec{\nabla}k_p\frac{\partial^2 \chi_r}{\partial k_p^2} \right]   =0.
\end{equation}
where $\hat{x}_p \equiv \vec{k}_p/k_p$ and, therefore, $\partial_{x_p}\mathcal{E}_p\equiv (\vec{k}_p.\vec{\nabla}\mathcal{E}_p)/k_p$.  Due to the consistency relation $\partial_t \vec{k}_p=-\vec{\nabla}\omega_p$, a transverse profile of the EPW frequency entails a transverse component in $\vec{k}_p$, usually directed towards the wave axis of propagation. Hence, in Eq.~(\ref{B3}), we do account for the self-focussing induced by wave front bowing, as discussed in Ref.~\cite{yin07}. Note, however, that making use of an adiabatic dispersion relation does not allow  for the phase modulation induced by the transverse dependence of the wave amplitude. Hence, we do not account for diffraction-like effects in Eq.~(\ref{B3}), and this point is discussed in detail in Appendix \ref{B}. 

Now, since the nonlinear frequency shift only depends on $\mathcal{E}_p$, $\partial_t \omega_p$ is directly proportional to $\partial_t \mathcal{E}_p$, and, actually,  $\mathcal{E}_p\partial^2_{\omega_p}\chi_r\partial_t \omega_p \ll \partial_{\omega_p}Ê\chi_r \partial_t \mathcal{E}_p$ because $\dwÊ\ll \omega_p$. Moreover, within the paraxial approximation, $k_p \approx k_{px}$, which yields  $\vec{\nabla}.\hat{x}_p \approx k_p^{-1}\vec{\nabla}_\bot.\vec{k}_p \equiv k_p^{-1}\left[\partial_y k_{py}+\partial_z k_{pz}\right]$. Finally, as shown in Ref.~\cite{PRL_10}, wherever SRS is effective, $\delta k_p \approx \dw/v_{gs}$, where $v_{gs}\approx c$ is the group velocity of the scattered wave, which makes  $\hat{x}_p.\vec{\nabla}k_p$ directly proportional to $\partial_{x_p}Ê\mathcal{E}_p$. Then, because $\dwÊ\ll \omega_p$ and the EPW phase velocity, $v_\phi$, is much less than $v_{gs}$, the term proportional to $\hat{x}_p.\vec{\nabla}k_p$ in Eq.~(\ref{B3}) is much smaller than that proportional to $\partial_{x_p}Ê\mathcal{E}_p$. Consequently, since in this paper we are mainly interested in the kinetic modeling of stimulated Raman scattering, we simplify Eq.~(\ref{B3}) in
\begin{equation}
\label{B5}
\frac{\partial \chi_r}{\partial \omega_p} \frac{\partial \mathcal{E}_p}{\partial t}-\frac{\partial \chi_r}{\partial k_p}\left[ \frac{\partial  \mathcal{E}_p} {\partial x_p}+\frac{\mathcal{E}_p}{2k_p} \vec{\nabla}_\bot.\vec{k}_p \right]=0.
\end{equation}

We now need to add to Eq.~(\ref{B5}) the terms accounting for collisionless dissipation. As discussed in Ref.~\cite{dissipation}, two different effects need to be allowed for, a Landau-like damping and the irreversible exchange of energy between the wave and the electrons induced by trapping. The expression we use for the collisionless damping rate follows from the theoretical developments of Ref.~\cite{benisti09}, which would yield, if all electrons had the same transverse velocity $\vec{v}_\bot$,
\begin{equation}
\label{nu1D}
\nu(\vec{v}_\bot)=\nu_{\text{lin}} \mathcal{H}\left[Y_{\text{3D}}\left(\vec{v}_\bot \right) \right],
\end{equation}
where $\nu_{\text{lin}}$ in the (linear) Landau damping rate~\cite{landau}, $Y_{\text{3D}}\equiv \int_{-\infty}^{t} \omega_B(x_\sslash-v_\phi t',\vec{x}_\bot-\vec{v}_\bot t',t') dt'$, $\omega_B \equiv \sqrt{e\mathcal{E}_pm/k_p}$ being the bounce frequency, and~$\mathcal{H}(Y_{\text{3D}})$~is a Heaviside-like function, $\mathcal{H}(Y_{\text{3D}})\approx 0$ if $Y_{\text{3D}} \agt 6$ and  $\mathcal{H}(Y_{\text{3D}})\approx 1$ if $Y_{\text{3D}} \alt 6$. 

As for the rate of dissipation induced by trapping, it is clearly proportional to the rate of electron trapping, which, if we neglect the small change in the EPW phase velocity,  is just proportional to the wave growth rate, calculated in the wave frame, and as seen by the electrons. For electrons with transverse velocity $\vec{v}_\bot$, the rate of dissipation induced by trapping is therefore proportional to $\partial_t\mathcal{E}_p+ v_\phi \partial_{x_p}\mathcal{E}_p+ \vec{v}_\bot.\vec{\nabla}_\bot\mathcal{E}_p$. Quite naturally, and as shown in Refs.~\cite{benisti07,dissipation}, in the envelope equation for the EPW the prefactor of the latter expression is $-\partial_{\omega_p} \chi_r^{\text{tr}}$, which is the contribution of the trapped electrons to $-\partial_{\omega_p} \chi_r$ . More precisely, $\chi_r^{\text{tr}}(\vec{r},\vec{v}_\bot,t)$ is the contribution to the real part of the electron susceptibility of those electrons with transverse velocity $\vec{v}_\bot$ which have been trapped by the EPW at time $t'\leq t$, and have completed at least one trapped orbit (i.e., $Y_{3D}Ê\agt 6$). Hence, $\chi_r^{\text{tr}}$ is a non local function of the plasma wave amplitude, which mainly depends on the maximum EPW amplitude experienced by the electrons with transverse velocity $\vec{v}_\bot$ (see Ref.~\cite{dissipation} for more details).

We therefore conclude that, when all electrons have the same transverse velocity, $\vec{v}_\bot$, allowing for collsionless dissipation leads to the following envelope equation for the plasma wave amplitude
\begin{equation}
\label{B5b}
\frac{\partial \chi_r^{\text{env}}}{\partial \omega_p}\left[\frac{\partial \mathcal{E}_p}{\partial t}+\nu(\vec{v}_\bot) \mathcal{E}_p \right]-\left[ \frac{\partial \chi_r}{\partial k_p}+v_\phi  \frac{Ê\partial \chi_r^{\text{tr}}}{\partial{\omega_p}} \right]  \frac{\partial  \mathcal{E}_p} {\partial x_p}-\frac{\partial \chi_r}{\partial k_p}\frac{\mathcal{E}_p}{2k_p} \vec{\nabla}_\bot.\vec{k}_p-\frac{Ê\partial \chi_r^{\text{tr}}}{\partial{\omega_p}} \vec{v}_\bot.\vec{\nabla}_\bot\mathcal{E}_p =0,
\end{equation}
where we have introduced $\partial_{\omega_p} \chi^{\text{env}}Ê\equiv \partial_{\omega_p} \chi_r-\partial_{\omega_p} \chi_r^{\text{tr}}$, whose expression may be found in Refs.~\cite{benisti07,brama}. When the transverse electron motion is not infinitely cold, one needs to average Eq.~(\ref{B5b}) over the distribution, $f(\vec{v}_\bot)$, of transverse velocities, which we approximate by the unperturbed one (assumed to be a Maxwellian), since the quiver transverse velocity induced by the laser field is much less than the thermal speed for the parameters investigated in this paper. Using the facts that $\partial_{\omega_p} \chi_r$ does not depend on $\vec{v}_\bot$, that $f(\vec{v}_\bot)$ is isotropic, and that $\partial_{k_p} \chi_r=(-2/k_p)\chi_r-v_\phi \partial_{\omega_p} \chi_r \approx (2/k_p)-v_\phi \partial_{\omega_p} \chi_r$ (see Ref.~\cite{vgroup} for details), averaging Eq.~(\ref{B5b}) over $\vec{v}_\bot$ yields
\begin{equation}
\label{B7}
\frac{\partial\chi^{\text{env}}_{\text{3D}}}{\partial \omega_p} \left[ \frac{\partial \mathcal{E}_p}{\partial t}+\vec{v}_g \vec{\nabla}\mathcal{E}_p+v_0 \frac{\mathcal{E}_p}{2k_p} \vec{\nabla}_\bot.\vec{k}_p+\nu_{3D} \mathcal{E}_p \right]=0,
\end{equation}
with
\begin{eqnarray}
\label{vgx}
v_{g_x}Ê&= &v_\phi-\frac{2}{k_p \partial_{\omega_p} \chi_{\text{3D}}^{\text{env}}Ê},\\
\label{vgy}
v_{g_{y,z}}& = &\left[v_\phi-\frac{2}  {k_p \partial_{\omega_p} \chi_{\text{3D}}^{\text{env}} } \right ]\frac{k_{y,z}}{k} +\frac{\int \partial_{\omega_p} \chi_r^{\text{env}} f(\vec{v}_\bot) v_{y,z} d\vec{v}_\bot}{  \partial_{\omega_p} \chi_{\text{3D}}^{\text{env}}  },\\
\label{v0}
v_0&=&v_\phi \frac{\partial_{\omega_p}\chi_r}{ \partial_{\omega_p} \chi_{\text{3D}}^{\text{env}}  }-\frac{2}{k_p \partial_{\omega_p} \chi_{\text{3D}}^{\text{env}}  },\\
\label{landau}
\nu_{3D}&=&\frac{\int f(\vec{v}_\bot) \nu(\vec{v}_\bot) \partial_{\omega_p} \chi^{\text{env}}(\vec{v}_\bot)d\vec{v}_\bot}{\partial_{\omega_p} \chi_{\text{3D}}^{\text{env}}  },
\end{eqnarray}
where $\partial_{\omega_p} \chi_{\text{3D}}^{\text{env}} \equiv \int f(\vec{v}_\bot) \partial_{\omega_p}Ê\chi_r^{\text{env}} d\vec{v}_\bot$. Note that, in order to derive $\vec{v}_g$, we made use of the approximation, $\vec{v}_\bot \approx v_y \hat{y}+v_z\hat{z}$. Note also that $\partial_{\omega_p}\chi_{\text{3D}}^{\text{env}}$ and $\partial_{\omega_p}\chi_r^{\text{env}}$ are non local functions of the EPW amplitude, which entails the dissipation of the electrostatic energy. More precisely, the expression we find for $v_{gx}$, and for the second term of Eq.~(\ref{vgy}) for $v_{gy}$ and $v_{gz}$, respectively lead to the longitudinal and transverse shrinking of the plasma wave packet in the strongly nonlinear regime, while maintaining the EPW amplitude constant along the characteristics, which automatically reduces the total electrostatic energy (see Ref.~\cite{dissipation} for details). As for the first term in Eq.~(\ref{vgy}), it allows for the self-focussing induced by wave front bowing. Moreover, it is quite clear from Eq.~(\ref{landau}) that we account for side-loss in our expression for the collisionless damping rate $\nu_{3D}$, and, from Eq.~(\ref{nu1D}), that  the latter would decrease less rapidly as a function of the EPW amplitude for a larger rate of side-loss (narrower wave packet or larger electron thermal velocity).

In order to complete our envelope equation for $\mathcal{E}_p$ we now need to account for the effect of the ponderomotive drive. From Eq.~(\ref{A11}) of Appendix \ref{A}, it is clear that, up to terms at second order in $k_y/k$, which we neglect, the ponderomotive force is the same as in 1-D. The right-hand side of the envelope equation for the EPW, accounting for the laser drive, is therefore exactly the same as the one derived in Ref.~\cite{brama}, so that, if we denote \begin{eqnarray}
\label{B8}
E_p&=&\mathcal{E}_p/2, \\
\label{B9}
E_s&=&(\mathcal{E}_s/2)e^{i\delta \varphi_p}\hat{y}.\hat{y}_s, \\
\label{B10}
E_l&=&(\mathcal{E}_l/2)\hat{y}.\hat{y}_l,\\
\label{B11}
\Gamma_p&=&\frac{ek_p}{m \omega_l \omega_s \partial_{\omega_p} \chi_{\text{3D}}^{\text{env}}},
\end{eqnarray}
we find
\begin{equation}
\label{B12}
\frac{\partial E_p}{\partial t}+\vec{v}_g \vec{\nabla} E_p+v_0 \frac{E_p}{2k_p} \vec{\nabla}_\bot.\vec{k}_p+\nu_{3D} E_p=\Gamma_p E_lE_s^*.
\end{equation}
Note that the solution of Eq.~(\ref{B12}) is not necessarily real as $E_p$ should be. Actually, enforcing the reality of $E_p$ would give rise to technical difficulties when numerically solving the EPW envelope equation, and would actually only make sense if we were able to perfectly calculate $\delta \varphi_p$, which is not the case. We therefore relax the constraint that $E_p$ needs to be real, and, although our procedure is not totally rigorous, it proved in Refs.~\cite{PRL_10,brama} to provide results in very good agreement with those of 1-D Vlasov simulations. This is because the only phase that has an impact of the growth of SRS is not that of the EPW itself, but the dephasing between the plasma wave and the laser drive, which we estimate by making use of the self-optimization ansatz detailed in Ref.~\cite{PRL_10}, and which is recalled in Section~\ref{III}.

\subsection{Equations for the laser and scattered waves}
The equations for the laser and scattered waves are derived from Maxwell laws by using the paraxial relation derived in Appendix \ref{A}, $E_x \approx (i/k)\partial_yE_y$, between the components of the electric field. More precisely, plugging Eq.~(\ref{A11}) into Eq.~(\ref{A3}) and making the usual approximations which consist in neglecting $E_z$, the space variations of $k$, the second derivatives in $x$ and $t$ as well as space derivatives of order larger than 3, yields
\begin{equation}
\label{B13}
\left[2ik\partial_x+2i(\omega/c^2) \partial_t+(\omega^2/c^2)-k^2Ê+\triangle_\bot \right]E_y=\mu_0(\partial_t-i\omega)j_y,
\end{equation}
where $\triangle_\bot \equiv \partial^2_{y}+\partial^2_{z}$. As is well known, $\mu_0(\partial_t-i\omega)j_y$ is the sum of the linear response to $E_y$, which yields $\omega_{pe}^2E_y$, and of the ``Raman'' term proportional to the density fluctuation, $\delta n e^{i(\varphi_p^{\text{lin}}+\delta \varphi_p)}+c.c.$, induced by the EPW (see Ref.~\cite{brama} for details). Since the relation $ik_p \mathcal{E}_p\approx-\delta n e /\varepsilon_0$ still holds, the ``Raman'' term is the same as in 1-D, so that one recovers the equations derived in Ref.~\cite{brama}, except for the additional term proportional to $\triangle_\bot E_y$. Hence, the envelope equation for the laser wave amplitude, $E_l$, is
\begin{equation}
\label{B14}
[\partial_t +v_{gl}\partial_x-i(v_{gl}/2k_l) \triangle_\bot ]E_l=\Gamma_l E_pE_s,
\end{equation}
where $v_{gl} \equiv k_lc^2/\omega_l$ and $\Gamma_l \equiv ek_p/2m\omega_s$. As regards the envelope equation for the scattered wave amplitude, $E_s$, we need to account for the extra phase $\delta \varphi_p$ in its definition, which yields,
\begin{equation}
\label{B16}
\left[\partial_t +v_{gs}\partial_{x_p}+i(\dw-v_{gs}\delta k_p)-i(v_{gs}/2k_s) \left(\triangle_\bot +i \vec{\nabla}_\bot.\vec{k}_p \right) \right]E_s=\Gamma_s E_lE_p^*,
\end{equation}
where $v_{gs} \equiv k_sc^2/\omega_s$ and $\Gamma_s \equiv ek_p/2m\omega_l$. 

\section{Comparisons with PIC simulation results}
  \label{III}
The previously derived envelope equations are solved numerically by our code, BRAMA, using the standard operator-splitting approach detailed in Ref.~\cite{brama}, with the additional hypothesis of self-optimization which consists in making the dephasing term $(\dw-v_{gs}\delta k_p)$ in Eq.~(\ref{B16}) as small as possible. More precisely, we calculate $\delta k_p$ from the consistency relation $\partial_t\delta \vec{k}_p=-\vec{\nabla}\dw$, and, when we find that $\delta k_p$ thus calculated may be larger than $\dw/v_{gs}$, we just  fix it at $\dw/v_{gs}$. This procedure proved to provide results in very good agreement with those of 1-D Vlasov simulations of SRS in Ref.~\cite{brama}.
 
Using BRAMA we aim at deriving the laser threshold intensity for SRS, as well as an order of magnitude for Raman reflectivity above threshold. In order to test the ability of our envelope code to do so, we compared its predictions against the 2-D and 3-D simulation results published by Yin \ea~in Refs.~\cite{yin08,yin09}, some of them we double-checked using our own PIC code CALDER \cite{calder}.

\begin{figure}
\centerline{\includegraphics[width=10cm]{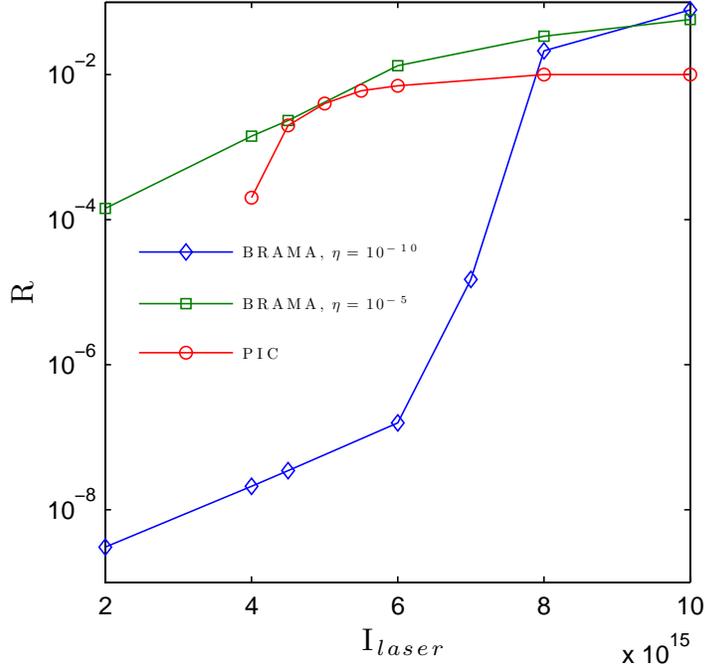}}
\caption{\label{f1} (Color online) Raman reflectivity, $R$, for a 2-D geometry, as obtained in Ref.~\cite{yin08} using PIC simulations (red circles), and derived from BRAMA simulations with $\eta=10^{-5}$ (green squares) or $\eta=10^{-10}$ (blue diamonds).}
\end{figure}

Let us start with the 2-D simulation results of Ref.~\cite{yin08}, corresponding to a plasma with electron temperature $T_e=700$ eV and electron density $n_e/n_{cr}=0.036$. From the left end of this plasma is injected a laser with wavelength $\lambda_l=0.527$ $\mu$m, which propagates along the $x$ direction and whose envelope at best focus, centered in simulation region, is a 2-D Gaussian whose intensity varies as $\exp (-z^2/w^2)$ with a waist $w=2.58$ $\mu$m. In BRAMA simulations is also launched, from the right end of the plasma, a seed laser which is not focussed, whose peak intensity, $I_s$, is related to that of the laser by $I_s=\eta I_{\text{laser}}$, and whose wavelength is $\lambda_s \approx 0.684$ $\mu$m (which maximizes the linear SRS growth rate for the chosen temperature and density). Then, the plasma wave generated by Raman is such that $k_p^{\text{lin}}\lambda_D \approx 0.34$, where $\lambda_D$ is the Debye length. The simulation domain has size 100 $\mu$m$\times14$ $\mu$m, which, in BRAMA simulations, we discretize using 400 points in the $x$ direction and 64 points in the $y$ direction. With BRAMA, it takes about 10 minutes on one processor to simulate 1 ps of laser-plasma interaction. As may be seen in Fig.~\ref{f1}, when $\eta=10^{-5}$, the same trend  is observed as regards the intensity dependence of Raman reflectivity calculated using either PIC or BRAMA simulations, which both predict that Raman scattering should not be very effective when $I_{\text{laser}} \alt 4\times 10^{15}$ W/cm$^2$ (although the onset of SRS is more marked in PIC than in BRAMA simulations). 
\begin{figure}
\centerline{\includegraphics[width=10cm]{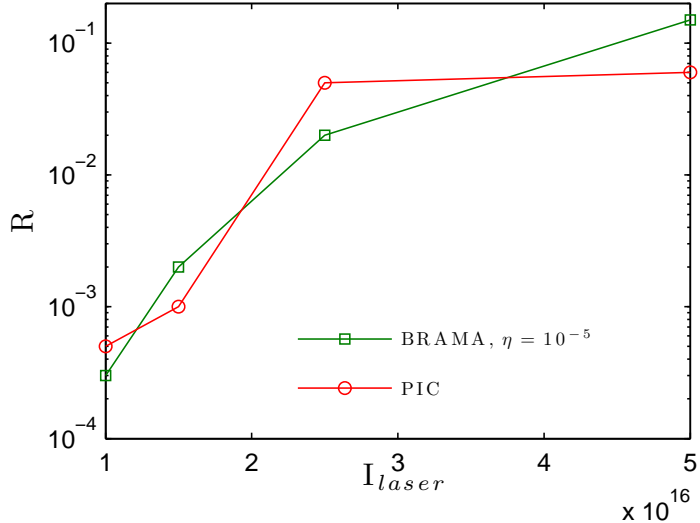}}
\caption{\label{f2} (Color online) Raman reflectivity, $R$, for a 3-D geometry, as obtained in Ref.~\cite{yin09} using PIC simulations (red circles), and from BRAMA simulations with $\eta=10^{-5}$ (green squares).}
\end{figure}

Now, one may wonder whether the good agreement found for the SRS threshold is  not just fortuitous, due to our particular choice for $\eta$, since in BRAMA Raman scattering results from the optical mixing of two counterpropagating lasers, while in PIC simulations it starts from the numeric noise. In order to address this issue, we strongly decreased $\eta$ (since we cannot greatly increase it without missing the linear regime), down to $\eta=10^{-10}$, which is way below the level of electromagnetic noise measured in our own PIC simulations with CALDER. As shown in Fig.~\ref{f1}, with such a low value of $\eta$ the threshold intensity for Raman scattering is $I_{th} \approx 7\times 10^{15}$ W/cm$^2$. Hence, the dependence of $I_{th}$ with respect to $\eta$ is very weak, a decrease of $\eta$ by 5 orders of magnitude leads to an increase in $I_{h}$ by a factor less than 2! This shows the little impact of the choice of $\eta$ on our results, and therefore confirms our ability to predict the SRS threshold using BRAMA. 

As regards Raman reflectivity above threshold, BRAMA results systematically overestimate the PIC ones, especially when $I_{\text{laser}}=10^{16}$ W/cm$^2$. For this particular intensity, we find that, as SRS reflectivity increases, the EPW experiences a strong self-focussing due to wave-front bowing, which should limit Raman scattering. More precisely, when $R \alt 1\%$, self-focussing is not yet effective, so that our modeling relying on the paraxial approximation should be valid, and we can therefore predict that, when  $I_{\text{laser}}=10^{16}$ W/cm$^2$, Raman reflectivity should reach at least 1\%. When $RÊ\approx6\%$, the EPW is strongly focussed, its the transverse extent is significantly less than the laser wavelength, and we have to stop the simulation. Indeed, such a result is clearly not physical and, although we are currently not able to prove it, it is most likely that Raman should saturate by some means (usually due to the growth of sidebands) before $R$ reaches 6\%. We therefore conclude that, when  $I_{\text{laser}}=10^{16}$ W/cm$^2$, SRS should saturate at a value of $R$ between 1\% and 6\% (and it is noteworthy that, in the PIC simulations of Ref~\cite{yin08}, the first maximum of SRS reflectivity is at $R \approx 6\%$  when $I_{\text{laser}}=10^{16}$ W/cm$^2$). Hence, for the time being, BRAMA simulations  cannot provide an extremely accurate estimate for the long time averaged Raman reflectivity, but can nevertheless predict what should be its order of magnitude, typically a few percents when  $I_{\text{laser}}=10^{16}$ W/cm$^2$, which is consistent the PIC results of Ref.~\cite{yin08} yielding $\langle R \rangle \approx 1.5\%$. It is also noteworthy that, as shown in Fig.~\ref{f1}, our results for $R$ above threshold do not depend on our choice for $\eta$. 

The same conclusions as before hold for the 3-D simulation results plotted in Fig.~\ref{f2}, which correspond to a plasma with electron temperature $T_e=4$ keV and electron density $n_e/n_{cr}=0.14$. The laser wavelength is  $\lambda_l=0.351$ $\mu$m, its waist is $w=1.4$ $\mu$m, and it is focussed in the center of the simulation box whose size is $35\times6\times6$ $\mu$m$^3$. In BRAMA simulations, the seed wavelength is $\lambda_s \approx 0.626$ $\mu$m (the corresponding EPW is such that $k_p^{\text{lin}}\lambda_D \approx 0.32$). The simulation domain is discretized using 100 points in the $x$ direction and 32 points in the $y$ direction. With BRAMA, it takes about 1 minute on 128 processors to simulate 1 ps of laser-plasma interaction. As may be seen in Fig.~\ref{f2}, and as for the previous 2-D simulations, we observe the same trend regarding the evolution of $R$ as a function of $I_{\text{laser}}$ using PIC or BRAMA simulations.  In particular, Raman reflectivity is expected to be very small when $I_{\text{laser}}<10^{16}$ W/cm$^2$, and to be of the order of a few percents when $I_{\text{laser}}=5\times10^{16}$ W/cm$^2$, although, for this particular intensity, the BRAMA prediction for $R$ overestimates that of the PIC simulation.

Using BRAMA, we actually reran all the simulations presented in Refs.~\cite{yin08,yin09}, and always found results similar to those shown in Figs.~\ref{f1} and \ref{f2}. We may therefore conclude that, both in 2-D and in 3-D, BRAMA is able to correctly estimate the SRS threshold and that, above threshold, it provides the correct order of magnitude for $R$. Therefore, using BRAMA one should be able to conclude about the effectiveness of Raman scattering, at least for a single laser speckle. 

\section{Conclusion}
\label{IV}
In this paper, we derived coupled envelope equations modeling stimulated Raman scattering in a multidimensional geometry, and allowing for such nonlinear kinetic effects as the frequency and wave number shift of the electron plasma wave, the nonlinear reduction of its collisionless damping rate, and the collisionless dissipation induced by trapping. 

Our theoretical predictions for the aforementioned nonlinear kinetic effects were shown to be in very good agreement with results from 1-D Vlasov simulations of SRS in Refs.~\cite{benisti08,benisti09,PRL_10}. We are therefore using here a nonlinear kinetic modeling of SRS that we previously checked very carefully, and that we merely generalized to allow for a multi-dimensional geometry so as to derive envelope equations which we solved using our envelope code BRAMA. In order to test BRAMA predictions as regards Raman reflectivity, we took advantage of the numerous 2-D and 3-D PIC results published by Yin~\ea~in Refs.~\cite{yin08,yin09}. When making comparisons which such PIC results, we had to be careful that, in BRAMA, we simulate the optical mixing of two counterpropagating lasers, while SRS starts from numerical noise in PIC simulations. In particular, we checked that the good agreement we found between the PIC and BRAMA predictions as regards the laser threshold intensity, $I_{th}$,  was not fortuitous, not due to the particular choice we made for the seed intensity, $I_s$, by showing the very weak dependence of $I_{th}$ with respect to $I_s$. 

Clearly, in BRAMA, we do not account for all possible nonlinear kinetic effects and, in particular, not for the growth of sidebands, although this is something we are currently working on, following the lines of Ref.~\cite{evangelos}. Consequently, we cannot make any claim regarding our ability to predict the long time averages of Raman reflectivity. We nevertheless interpret the strong EPW self-focussing we observe in BRAMA simulations as a natural limit for Raman growth, which allows us to predict an order of magnitude for SRS reflectivity above threshold, which appears to be consistent with that of the PIC simulations. Hence, although it is clear that the present paper is certainly not the last word on the nonlinear kinetic modeling of SRS, we may nevertheless conclude that, in its present state, BRAMA appears to be a very powerful tool to predict Raman effectiveness, at least for a single laser speckle.

\appendix
\section{Discussion on the nearly electrostatic, or electromagnetic, nature of the waves.}
\label{A}
\setcounter{equation}{0}
\newcounter{app}
\setcounter{app}{1}

Let us consider an electric field, $\vec{\mathcal{E}}$, which we write 
\begin{equation}
\label{A1}
\vec{\mathcal{E}} = \left[E_x \hat{x}+E_y \hat{y} +E_z \hat{z}Ê\right]e^{i(kx-\omega t)}+c.c.
\end{equation}
From Maxwell equations, we know that 
\begin{equation}
\label{A2}
\vec{\nabla}^2 \vec{\mathcal{E}}-c^{-2}\partial^2_t \vec{\mathcal{E}}=\mu_0 \partial_t \vec{j}+\vec{\nabla}Ê[\vec{\nabla}.\vec{\mathcal{E}}Ê],
\end{equation}
where $\vec{j}$ is the current density. Projected on the $x$, $y$, and $z$ directions, Eq.~(\ref{A2}) yields
\begin{eqnarray}
\nonumber
\left[\partial^2_y+\partial^2_z +c^{-2}\left( \omega^2-\partial^2_t+2i\omega \partial_t \right)\right]E_x&=&\mu_0 (\partial_t -i\omega) j_x +(\partial^2_{xy}+ik \partial_y)E_y\\
\label{A2b}
&&+(\partial^2_{xz}+ik \partial_z)E_z \\
\nonumber
\left[\partial^2_x+\partial^2_{z,y} +2ik\partial_x-k^2+c^{-2}\left( \omega^2-\partial^2_t+2i\omega \partial_t \right)\right]E_{y,z}&=&\mu_0 (\partial_t -i\omega) j_{y,z}+ik\partial_{y,z}E_x\\
\label{A3}
&&+\partial^2_{xy,z}E_x+\partial^2_{yz}E_{z,y}. 
\end{eqnarray}
\subsection{Nearly electrostatic nature of the plasma wave}
The nearly electrostatic nature of the plasma wave is best viewed by writing Eq.~(\ref{A2}) in Fourier space, which yields, along the $y$ or $z$ direction,
\begin{equation}
\label{A8}
\left[w^2/c^2-\kappa^2 \right] \tilde{\mathcal{E}}_{y,z}+i\mu_0 w \tilde{j}_{y,z}=-\kappa_{y,z}(\kappa_x\tilde{\mathcal{E}}_x+\kappa_y\tilde{\mathcal{E}}_y+\kappa_z\tilde{\mathcal{E}}_z).
\end{equation}
For a freely propagating wave, $i\mu_0 w \tilde{j}_{y,z}=-(w^2/c^2)\tilde{\mathcal{E}}_{y,z}$, while for a driven wave one needs to account in $i\mu_0 w \tilde{j}_{y}$ for the additional term $-(w^2/c^2)\tilde{\mathcal{E}}_{dy}$, induced by the $y$ component of the driving field. Now, because the electron thermal velocity is much less than the speed of light, $w^2/c^2 \ll \kappa^2\approx \kappa^2_x$. Moreover, from Eq.~(\ref{A10}), $\tilde{\mathcal{E}}_{dy}Ê\approx (\kappa_y/\kappa_x) \tilde{\mathcal{E}}_{dx}$, while $\vert \tilde{\mathcal{E}}_{dx} \vert \ll \vert \tilde{\mathcal{E}}_x \vert$. Hence, we conclude that $(w^2/c^2)\vert \tilde{\mathcal{E}}_{dy} \vert$ is much less than $\vert \kappa_y \kappa_x \mathcal{\tilde{E}}_x \vert$, and is thus negligible in Eq.~(\ref{A8}), whose left-hand side may therefore be approximated by $-\kappa^2 \tilde{\mathcal{E}}_{y,z} \approx -\kappa_x^2 \tilde{\mathcal{E}}_{y,z}$. Moreover, since we only aim at modeling the nearly paraxial propagation of the plasma wave, which is essentially polarized along the $x$-direction, the right-hand side of Eq.~(\ref{A8}) is close to $-\kappa_{y,z}\kappa_x \tilde{\mathcal{E}}_x$. Hence,
\begin{equation}
\label{A9}
\tilde{\mathcal{E}}_{y,z}\approx (\kappa_{y,z}\kappa_x/\kappa^2)\tilde{\mathcal{E}}_xÊ\approx (\kappa_{y,z}/\kappa)\tilde{\mathcal{E}}_x.
\end{equation}
The significant Fourier modes of the EPW electric field are nearly electrostatic, and, therefore, so is the total field. 

\subsection{Longitudinal component of the laser and scattered wave electric fields}
We now assume that $\vec{E}$ is essentially polarized along the $y$ direction, which is the case for the laser and scattered waves. Then, at lowest order in the field variations, Eq.~(\ref{A2b}) is
\begin{equation}
\label{A10}
(\omega^2/c^2)E_x=-i\mu_0\omega j_x+ik \partial_y E_y.
\end{equation}
Since $-i\omega j_x \approx \omega_{pe}^2 E_x$ (if we neglect the  density modulation), and provided that $\omega$ and $k$ are chosen so that $\omega^2=\omega_{pe}^2+k^2c^2$, Eq.~(\ref{A10}) yields
\begin{equation}
\label{A11}
E_x \approx (i/k) \partial_y E_y.
\end{equation}
As for $E_z$, from Eq.~(\ref{A3}) we would find that it remains at very low levels, and we henceforth neglect it.

\section{Diffraction terms for the plasma wave}
\label{B}
\setcounter{equation}{0}
\setcounter{app}{2}

Diffraction-like effects for the plasma wave are only addressed in the linear regime, where we slightly change our notations and write the EPW electric field $\vec{E}_{\text{EPW}} \equiv \vec{E}_pe ^{i(k_0x-\omega_0t)}+c.c.$, and the electron charge density $\rho=\rho_0e ^{i(k_0x-\omega_0t)}+c.c.$ Then, if we introduce 
\begin{equation}
\label{B17}
\chi=\frac{i\rho_0}{\varepsilon_0k_0(E_{px}+iE_{dx})},
\end{equation}
where, using the same notations as in Section~\ref{II}, $E_{dx}\equiv ek_0E_lE_s^*/(m\omega_l\omega_s)$ is the $x$-component of the driving field, Gauss law yields,
\begin{equation}
\label{B18}
(1+\chi)E_{px}-\frac{i}{k_0} \vec{\nabla}.\vec{E}_p=-i\chi E_{dx}.
\end{equation}
In order to calculate $\chi$, we use a Fourier representation of the fields and write
\begin{equation}
\label{B19}
\rho_0  \equiv -i\varepsilon_0 \int \xi(k,\omega)\vec{k}.\vec{\tilde{E}}e^{i(\vec{k}.\vec{r}-\omega t)} d\vec{k} d\omega,
\end{equation}
where $\vec{E}$ is the sum of the EPW and driving fields and, since each Fourier mode of the EPW field is nearly electrostatic, $\xi$ only depends on the wave number $\vec{k}$ through its modulus, $k$, as indicated in the integral of Eq.~(\ref{B19}). This integral is approximated by making use of a Taylor expansion about $\vec{k}=k_0 \hat{x}$ and $\omega=\omega_0$, at first order in $\delta k_x$, $\delta \omega$, $\delta k_y^2$ and $\delta k_z^2$, 
\begin{equation}
\label{B20}
\xi(k,\omega)\vec{k}.\vec{\tilde{E}} \approx \xi(k_0,\omega_0) \left[k_0 \tilde{E}_{x}+\delta \vec{k}.\vec{\tilde{E}} \right ]+k_0 \tilde{E}_{x}\left[\frac{\partial x}{\partial k_0} \left(\delta k_x+\frac{\delta k_y^2}{2k_0}+\frac{\delta k_z^2}{2k_0} \right) +\frac{\partial \xi}{\partial \omega}\delta \omega \right],
\end{equation}
which yields
\begin{equation}
\label{B21}
\chi \approx \xi(k_0,\omega_0) \left[1-\frac{i}{k_0}\vec{\nabla}.\vec{E}_p\right]-i\frac{\partial \xi}{\partial k_0} \frac{\partial E_{px}}{\partial x}+i\frac{\partial \xi}{\partial \omega_0}\frac{\partial E_{px}}{\partial t}-\frac{1}{2k_0} \frac{\partial \xi}{\partial k_0} \triangle_\bot E_{px},
\end{equation}
where we made use of the approximation $E_x^{-1} \partial_u \vec{E}Ê\approx E_{px}^{-1}Ê\partial_u \vec{E}_p$ for $u=x,y,z,t$. This approximation is vindicated in the regime of strong damping because, in this regime, $\vec{E}_p$ and $\vec{E_d}$ are nearly proportional, while, when the damping is small, $E_d \ll E_p$ and the approximation still holds. Plugging Eq.~(\ref{B21}) into Eq.~(\ref{B18}), we find
\begin{equation}
\label{B22}
\left[ 1+ \xi(k_0,\omega_0)\right]    \left(E_p-\frac{i}{k_0}\vec{\nabla}.\vec{E}_p\right)-i\frac{\partial \xi}{\partial k_0} \frac{\partial E_{px}}{\partial x}+i\frac{\partial \xi}{\partial \omega_0}\frac{\partial E_{px}}{\partial t}-\frac{1}{2k_0} \frac{\partial \xi}{\partial k_0} \triangle_\bot E_{px}=-i\chi E_{dx}.
\end{equation}
When the Landau damping rate, $\nu_{\text{lin}}$, is small compared to the plasma frequency, $\xi(k_0,\omega_0) \approx \xi_r(k_0,\omega_0)+i\partial_{\omega_0}Ê\xi_r \nu_{\text{lin}}$, where $\xi_r$ is calculated by making use of the adiabatic approximation. Hence, $\xi_r$ assumes the same values as what we previously used for $\chi_r$ in the variational approach of Section~\ref{II}. Thus, if we choose $k_0$ and $\omega_0$ such that $1+\xi_r(k_0,\omega_0)=0$, Eq.~(\ref{B22}) becomes
\begin{equation}
\label{B23}
\left[ \partial_t +v_{g0}\partial_x-i(v_{g0}/2k_0)  \triangle_\bot +\nu_{\text{lin}} \right] E_{px}=\Gamma_pE_lE_s^*,
\end{equation}
where $v_{g0}Ê\equiv -\partial_{k_0}\xi_r/\partial_{\omega_0} \xi_r$ and where, as usual, we calculated the right-hand side of the envelope equation at lowest order in the variations of the field amplitudes, so that $\Gamma_p$ is given by Eq.~(\ref{B11}). Is is quite clear that Eq.~(\ref{B23}) is very similar to Eq.~({\ref{B14}) for the laser wave amplitude. Now, because $v_{g0}Ê\ll v_{gl,s}$, it is also quite clear that diffraction is much less effective for the plasma wave than for the laser or scattered wave. Since, as regards SRS, only the \textit{relative}~phase between the plasma wave amplitude and that of the ponderomotive drive,  and the space overlap of these two fields, matter, we conclude that accounting for a diffraction term in the EPW envelope equation is not essential. We actually checked this by including a linear diffraction term in the EPW envelope equation and found no significant difference in our results for Raman reflectivity. 

Now, if we denote $E_{px}Ê\equiv E_0e^{i\varphi_x}$, with $E_0 \equiv \vert E_{px}Ê\vert$, then, provided that the plasma wave and the laser drive are perfectly in phase, Eq.~(\ref{B23}) yields
\begin{equation}
\label{B24}
\frac{\partial E_0}{\partial t}+v_{g0} (\vec{k}_p/k_0) \vec{\nabla} E_0+v_{g0} \frac{E_0}{2k_0} \vec{\nabla}_\bot.\vec{k}_p+\nu_{\text{lin}} E_0=\Gamma_p \vert E_lE_s^*Ê\vert,
\end{equation}
where $\vec{k}_p \equiv k_0 \hat{x}+\vec{\nabla} \varphi_x$. Eq.~(\ref{B24}) is just the linear limit of Eq.~(\ref{B12}), except that it is written for the $x$-component of the field amplitude and not for its norm (actually, replacing $E_0$ with $E_p$ only yields terms of higher order than that at which the envelope equations were derived). We therefore conclude that Eq.~(\ref{B12}) derived in Section \ref{II} only misses the phase modulation due to the transverse variation of the field amplitude. More precisely, still when the EPW and the laser drive are perfectly in phase, Eq.~(\ref{B23}) also yields
\begin{equation}
\label{B25}
\partial_k \xi_r \left[\delta k_x+(\delta k_y^2+\delta k_z^2)E_0/2k_0-(\triangle_\bot E_0)/2k_0  \right]+\partial_\omega \delta \omega=0,
\end{equation}
where $\delta \vec{k} \equiv \vec{\nabla}Ê\varphi_x$ and $\delta \omega \equiv -\partial_t \varphi_x$, which shows that the central wave number and frequency of the EPW are, respectively, $\vec{k}_p \equiv k_0\hat{x}+\delta \vec{k}$ and $\omega_p \equiv \omega_0+\delta \omega$ such that
\begin{equation}
\label{B26}
\left[1+\xi_r \left( \vert \vec{k}_p \vert, \omega_p\right) \right]E_0=\partial_k \xi_r(\triangle_\bot E_0)/2k_0.
\end{equation}
The former equation makes explicit how the EPW dispersion relation is affected by the transverse variations of the EPW amplitude. However, in the nonlinear regime, the transverse phase modulation of the plasma wave is mainly induced by the adiabatic frequency shift, $\dw$. Indeed, as discussed in Ref.~\cite{benisti07}, $\dw$ may by calculated by making use of the adiabatic approximation, and therefore at 0-order in the transverse variations of the EPW, provided that $v_{th}/l_\bot \alt \omega_{pe}/20$, where $l_\bot$ is the typical transverse gradient length of the EPW and $v_{th}$ the electron thermal temperature. For the parameters used in this paper, $v_{th}/l_\bot \alt \omega_{pe}/20$ if $l_\bot \agt \lambda_l$, where $\lambda_l$ is the laser wavelength. 


\end {document}